\documentclass[aps,prb,twocolumn]{revtex4}
\usepackage{graphicx}

\begin{document}

\newcommand{\beq}{\begin{equation}}
\newcommand{\eeq}{\end{equation}}
\newcommand{\bqa}{\begin{eqnarray}}
\newcommand{\eqa}{\end{eqnarray}}
\newcommand{\nn}{\nonumber}
\newcommand{\nl}{\nn \\ &&}
\newcommand{\dg}{^\dagger}
\newcommand{\rt}[1]{\sqrt{#1}\,}
\newcommand{\erf}[1]{Eq.~(\ref{#1})}
\newcommand{\Erf}[1]{Equation~(\ref{#1})}
\newcommand{\smallfrac}[2]{\mbox{$\frac{#1}{#2}$}}
\newcommand{\bra}[1]{\left\langle{#1}\right|}
\newcommand{\ket}[1]{\left|{#1}\right\rangle}
\newcommand{\ip}[2]{\left\langle{#1}\right|\left.{#2}\right\rangle}
\newcommand{\sch}{Schr\"odinger }
\newcommand{\schs}{Schr\"odinger's }
\newcommand{\hei}{Heisenberg }
\newcommand{\heis}{Heisenberg's }
\newcommand{\half}{\smallfrac{1}{2}}
\newcommand{\bl}{{\bigl(}}
\newcommand{\br}{{\bigr)}}
\newcommand{\ito}{It\^o }
\newcommand{\sq}[1]{\left[ {#1} \right]}
\newcommand{\cu}[1]{\left\{ {#1} \right\}}
\newcommand{\ro}[1]{\left( {#1} \right)}
\newcommand{\an}[1]{\left\langle{#1}\right\rangle}
\newcommand{\implies}{\Longrightarrow}
\newcommand{\tr}[1]{{\rm Tr}\sq{ {#1} }}
\newcommand{\st}[1]{\left| {#1} \right|}

\newcommand{\red}[1]{\color{red}{#1}\color{black}}
\newcommand{\blu}[1]{\color{blue}{#1}\color{black}}

\title{An Entangled Web of Crime: Bell's Theorem as a Short Story}

\author{Kurt Jacobs}
\author{Howard M. Wiseman}
\affiliation{Centre for Quantum Computer Technology, Center for Quantum Dynamics, School of Science, Griffith University, Nathan 4111, Australia}

\begin{abstract}

Non-locality of the type first elucidated by Bell in 1964 is a difficult
concept to explain to non-specialists and undergraduates. Here we attempt this
by showing how such non-locality can be used to solve a problem in which someone
might find themselves as the result of a collection of normal, even if
somewhat unlikely, events. Our story is told in the style of a Sherlock Holmes
mystery, and is based on Mermin's formulation of the ``paradoxical"
illustration of quantum non-locality discovered by Greenberger, Horne and
Zeilinger.

\end{abstract} 

\maketitle

\section*{Preamble} 

With the discovery of Bell's theorem in 1964,\cite{Bel64} and the experiments
it prompted over the next two decades,\cite{Aspect} an astonishing fact about
the nature of the universe was revealed: it is non-local. That is, that certain
things which happen in the universe can only be explained if there is 
instantaneous action-at-a-distance, although this cannot be used to 
communication instantaneously. The exact nature of the non-locality is thus very
subtle  and not as easy to explain to a general readership as
other key insights about the physical universe such as the invariance of the
speed of light, or even Heisenberg's uncertainty principle. 

In this article we illustrate the weird non-locality of quantum mechanics, 
which is the import of Bell's theorem, using the literary device of a detective
story.  While a few previous articles have been written with a similar purpose
--- that is, to explain quantum non-locality using everyday
settings\cite{Jack,Price,Penrose,gameshow,cakes,Avarind} --- in none of these
was the use of the non-locality required to solve a problem with which  someone
might be faced as the result of everyday, if rather  coincidental, events. 
This article was motivated by a desire to construct an example of such a 
 situation. We feel that our story should be of use in engaging
students,  and with this in mind we have included a number of exercises
throughout the  story, the answers to which are given at the end of the
article.

Like Ref. \cite{gameshow} we base our story around the example of nonlocality, 
described in this form by Mermin, which uses three-party GHZ entanglement.
\cite{GHZ,Mermin90,GHSZ,CRB}  Reference\cite{Bra04} gives a comprehensive
review of nonlocality of this sort, which the authors call ``quantum
pseudo-telepathy". In this the authors show that the GHZ example is the
simplest possible in the sense of requiring the smallest Hilbert-space
dimension.  Other examples from Ref.~\cite{Bra04} may however be quicker to
explain, in  particular one by Aravind~\cite{Ara02}. We have been unable to
construct a  compelling story around this  example, but we encourage the reader
to try.  A much broader review of ``strange correlations, paradoxes and
theorems" in  quantum mechanics may be found in Ref. \cite{Lal01}.

We were prompted to write this article as a reaction to the (throwaway)
statement by Mermin that ``the action at a distance [in Bell's theorem] is
entirely useless.''\cite{Mermin} As the story shows, it is certainly not
useless. Bell-type  non-locality does not break Einstein's no-signaling
condition, but that does not make it  any less real. Of course quantum
non-locality is known to be potentially useful for practical tasks such as 
scheduling with a minimum of classical communication,\cite{schedule} but a) the
protocols  for these tasks are far more complicated, and b) the effect appears
less dramatic than for the one we describe.  

There {\em are} simple tasks, such as  quantum teleportation, or dense coding,
which rely upon quantum entanglement and are usually understood to involve
quantum non-locality. However, for a non-specialist to appreciate any of the
weirdness in these examples he or she must first understand a substantial
amount of quantum mechanics.\cite{footnote} Moreover, recent 
studies\cite{HardyTel,Spe04} show that these (and many other) tasks in quantum
information can be simulated in a quantum-like theory which is completely
local. It seems that Bell's theorem is still the best way of illustrating the
non-locality of the world.

The story below is told in the style of a Sherlock Holmes mystery. The narrator
is Mr.\ Doyle, and the protagonist is Dr.\ Bell. As is now well-known, the 
chief inspiration for Arthur Conan Doyle's most famous literary creation,  Sherlock Holmes,
was a Dr.\  Bell who lectured Doyle at Edinburgh University
Medical School. A fictionalized version of their relationship has been told in
a number of recent novels,\cite{Engel,Pirie} in which Mr.\ Doyle plays Watson
to Dr.\ Bell's Holmes,  and we model our story loosely on that pattern.

\section*{The Case of the Two-Colour Gang}

It was a late afternoon early in October when first I found myself outside the
door of number 8 Hilbert Place, a rather nondescript two-storey house in a small
street just south of the city center. While it was mid autumn, the sky was
clear, and the afternoon warm as the sun's rays lingered on the trees and
grounds of number 8. The bright weather was in some contrast to my mood
however, as I was weighed down with a problem which had been occupying my mind
for some days. I rang the doorbell, and as I did so my thoughts wandered back
over the events of the last few weeks. 

I made my living as a barrister, and my private legal practice was doing very
well. I enjoyed my work, and had been attracting cases both of increasing
interest and importance. A few months before I had been lucky enough to land a 
defense case which was very much in the public eye --- certainly it was  the
highest profile case of any with which I had then been associated.  As events
would turn out, it would also be one of my greatest triumphs, for  (despite the
evidence against them) the case against my clients was dropped. Now,  many
years later, the details of the case can be told for the first time. 

The preceding summer had seen a number of
break-ins at the Museum of Semi-Classical Art. This had been the cause of
considerable concern as the museum was due to host the exhibition {\em Local
Realism}, a collection of very valuable works by artists of the
world-renowned realist school that arose in our city. In view of this
the curator had increased security by placing four guards in the newly
built Isosceles wing, which was to house the collection. This precaution
was indeed prudent, for a mere three days after the exhibition  opened a
very daring robbery was attempted. Three robbers somehow defeated the
perimeter alarms and broke into the museum at midnight. They split up and
dashed through the corridors of the Isosceles wing, each aiming to grab a
particularly valuable piece of art. Their plans were foiled by the
guards, however, who spotted them and raised the alarm. While this
prevented  the robbers from carrying off any of the art, the guards did
not manage to catch them.

Fortunately for the police, the descriptions given by the guards fitted three
well known criminals, and the next day they made a dawn raid on their home.
There the police found further evidence linking the three to the attempted
robbery, and they were subsequently arrested and charged. I had myself only
just put down the evening paper, where I learnt of the arrest, when quite out
of the blue I received a call from the three men in custody. Having no other
matter of importance on hand at the time, I agreed to represent them, but
certainly had no idea what a curious turn the case would take. 

My reverie on the doorstep of number 8 Hilbert Place was broken by footsteps 
in the hall. The door opened to reveal a
young woman with a pleasant face and a bright smile. She was dressed in casual
clothes, with a loose-fitting pullover and blue jeans. While one would not
immediately associate such attire with that of a consultant to a prestigious
legal firm, it was only when I noticed the fluffy Bugs Bunny slippers that I
wondered for an instant if I had indeed knocked at the right door.  

`You must be Mr.\ Doyle', said the young woman. `I'm Alice Bell. Do come in.' 

`Thank you', I said. `It was good of you to see me at such short notice' 

`Not at all Mr.\ Doyle. As you probably know I do most of my consulting for the
legal firm Greenberger-Horne-Zeilinger, but they have offered few cases of late
which exhibit those singular features so necessary if the problem is to provide
any real interest for me. I assure you the debt will be more than repaid if
your case is of sufficient curiosity'.

`It certainly seems so to me, I must admit', I replied.

`Excellent', said Dr. Bell, `Cup of tea?' 

I realized that I was indeed thirsty, and as she handed me a steaming cup and
took a plate of cookies from the sideboard, I realized that I was also
quite hungry, having not eaten since breakfast. She then led me up some
stairs to a pleasantly furnished office --- along with the mandatory desk and
laptop it housed two comfortable and somewhat weatherbeaten leather chairs and
a small coffee table. Shelves lined two adjacent walls, and while many
were filled with books, others contained jars of various shapes and sizes
which I assumed at first to contain chemicals, although later inspection
revealed a much greater and more unusual variety of contents. The third
wall was covered by a large and detailed map of the world, with the final
wall devoted almost completely to a huge window, affording a good view of
the city center, including a hint of the harbor and northern hills
beyond. To the side of the desk was what seemed to be some kind of
electronic apparatus, but apart from the brand name `Cryptolightning'
which was written on the side there was no indication as to its function. 

Placing the cookies on the coffee table she sat down and motioned me to take
the other chair. I did so, and as I helped myself to a cookie she slid a little
further into her chair, and placing the tips of her fingers together said with
a slight smile and an unmistakable air of anticipation `So Mr.\ Doyle, what is
it that brings you here?'

`You have heard of the break-in at the Museum of Semi-Classical Art?', I asked.

`It is hardly possible not to have', she said `You may safely assume that I am
familiar with all that has been in the papers, but no more'.

`Then I shall begin straight in with the details' I said. `As you know from the
papers, my clients were discovered with a number of body suits of the type worn
by cat burglars. While some of these were a single color, being red or green,
the others were more unusual in that they were green on the front and red on
the back, or vice versa. Obviously the prosecution wanted to form as strong a
link as possible between these suits and those that the robbers were wearing,
so naturally I cross-examined the guards very carefully on this point.'

`Naturally', murmured Dr.\ Bell as I paused to take another bite.

`Now, one must understand' I continued, `that the lighting in the Gallery was
rather odd, changing in color and intensity from place to place in accordance
with the artwork. As a result the guards were not able to discern completely
the colors of the robbers' clothes. However, the first three asserted that 
they had had a clear view of one of the robbers, and that he was wearing a
red suit, but none of them could remember which robber it was. Thus it may
even have been a different robber in each case. In addition, they were sure
that the other two robbers were wearing the same color, but they could not be
sure what color it was under the  lighting conditions. The testimony of the
fourth and final guard was only a little different. He asserted that one of the
robbers was wearing a green suit, and that the other two were wearing the same
(although again unknown) color.' 

`So can we sum up the guards' statements by saying that the first three  guards
saw an odd number of robbers wearing red, and the fourth saw an even number
wearing red?', asked Dr.\ Bell.

`Most artfully put,' I said. `Now while it was not possible to conclude much 
from these statements alone, further evidence was provided by the infra-red 
security cameras, which showed clearly the  paths taken by the robbers as they 
ran through the gallery. In addition, the guards  made definite statements as 
to their respective locations when they saw the robbers.  In the light of this 
I was able to find an inconsistency in the guards' testimony.' 

\begin{figure}
\includegraphics[width=1.0\hsize]{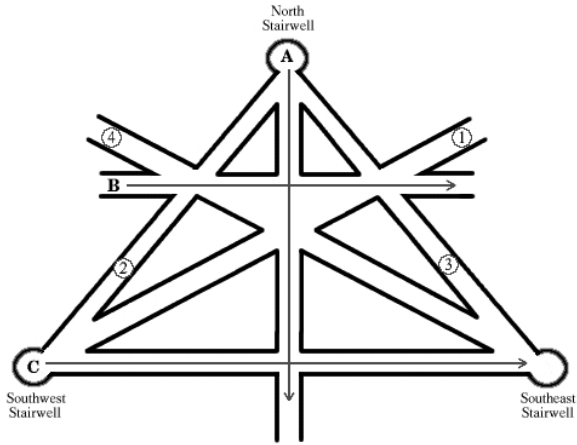}
\caption{Here is shown Mr. Doyle's map of the Isosceles wing of the Museum for
Semi-Classical art, with the positions of the four guards (marked by the
numbered circles), and the passage of the robbers A, B and C through the
gallery as recorded by the infra-red security cameras.}
\label{map}
\end{figure}

`I have here a map of the Gallery on which I have indicated the positions of
the guards and the paths taken by the robbers.' I drew the map from my
briefcase, and handing it to Dr.\ Bell, added `I have labeled the guards with
the numbers 1 to 4, and the robbers with the letters A B and C' (Note: Mr.\
Doyle's map is reproduced in Fig.~\ref{map})

I paused for a few moments, allowing Dr.\ Bell the chance to take in the map. 

`So each of the Guards saw only the back or the front of each robber, but not
both?', she asked, looking up from the map.

`Yes, indeed', I replied, most impressed by her perspicacity. `Thus the
testimony of each guard only refers to either the back or the front of each of
the robbers. From the map we know, for example, that the first guard saw the
back of robber A but the fronts of robbers B and C.  The statements of the
guards therefore refer not to three, but to six different things, being the two
sides of each of the three robbers. I found it convenient to summarize which
sides were seen by the various guards in a table.', and fishing the table out
of my briefcase I handed it to Dr.\ Bell. (Note: Mr.\ Doyle's table is
reproduced here in Table~\ref{table})~\cite{footnote2} 

\begin{table}
\begin{tabular}{|l@{\hspace{0.2cm}}|c@{\hspace{0.2cm}}|c@{\hspace{0.2cm}}|c@{\hspace{0.2cm}}|}
   \hline
      & {\bf Robber A} & {\bf Robber B} & {\bf Robber C} \\ 
   \hline 
   {\bf Guard 1} & Back & Front & Front \\
   {\bf Guard 2} & Front  & Back & Front \\
   {\bf Guard 3} & Front  & Front & Back \\ 
   {\bf Guard 4} & Back & Back  & Back \\                 
   \hline
\end{tabular}
\caption{Each of the guards saw either the back or the front of each of the robbers. Mr.\ Doyle's table reproduced here shows which of the two it was for each guard and each robber.}
\label{table}
\end{table}

`From this table, and the statements of the guards,' I went on,  `I was able to
show that while any three of the guards' claims are consistent, all four are not
--- one of them must be lying, or at the least mistaken.' In response to the
Doctor's raised eyebrows I proceeded to explain my reasoning, with which I must
say I was rather pleased.  

\vspace{3mm}
\hspace{-3.5mm}{\em Exercise 1:} Reproduce Mr.\ Doyle's argument. 
\vspace{3mm}

`After I had presented the argument in court,' I continued, `the prosecution
asked for a private word. It turned out that the police suspected that one of
the guards was working for the gang that organized the break-in, but didn't
know which one it was. If this was the case, then the guard in question was
almost certainly away from his post turning off the perimeter alarms when the
robbers broke in, and this would mean that he would have to have fabricated
his  evidence. If the police could find out which guard was lying, it would
give  them a new lead to the mastermind behind the robberies.' 

`So the prosecution offered me a deal. They said that if my clients would tell
them which guard was lying, they would drop the case against them. Now this
deal is indeed attractive, because there is considerable evidence against my
clients, and I certainly cannot guarantee a victory. However, my clients deny
any involvement in the attempted robbery, and if they are telling the truth
then they do not have the information the police want. And if (heaven forbid) 
they {\em are} members of the crime ring, it would be unwise for them to aid
the police; if the crime boss discovered that they led the police to him then
my clients would be better off in jail.'

`I thought at first that we might be able simply to make up a story as to what
the robbers were wearing, so as to accept the offer and satisfy the
prosecution. However, this fails for two reasons. The first is that the police
informed me that one of the guards is an undercover officer who's testimony is
beyond doubt. Thus, if our story conflicts with his testimony they will know we
are lying. The second reason is that, if by chance our story incriminates the
guard working for the crime ring, my clients could be in worse trouble.'

`I had virtually decided that I would have no option but to reject the offer.
However, I mentioned the problem last night to a colleague of mine who works
for GHZ, and she was adamant that before giving up I should come and see you
--- so there you have it.' Having finished my exposition I sat back and drained
my cup. 

Dr.\ Bell was silent for a few minutes, apparently lost in thought. At last she
said, `Your situation is indeed an interesting one, Mr.\ Doyle. Let us consider
what would happen if we let the police ask just one question of each of your
clients, being {\em either} what color his suit was on the front, {\em or}
what color it was on the back.'

I picked up my table again, which Dr. Bell had placed on her coffee table, and 
examined it as she continued. 

`I think you will find that in this case the police will only be able to test
one of the guards' statements, rather than all four, but they will be able to 
test any one of the statements by choosing which question they ask each of your
clients'. 

`Yes, that seems to be right', I said, after studying the table. However, I 
don't see how this would help us. It is true that if my clients were to know
which question each was to be asked, they would know which guard's statement
was being tested, and thus what to answer so as to confirm that statement. That
way they could be sure neither to contradict the undercover policeman, nor to
unwittingly finger the crooked guard. However, the police will surely demand to
question each of my clients separately. Moreover, in a case as important as
this they will no doubt place each of them in a sealed room, in different
buildings, so as to prevent absolutely any form of communication between them.
Thus none of them will know which questions the others are being asked. Without
that information they won't know what to answer.'

\vspace{3mm}
\hspace{-3.5mm}{\em Exercise 2:} (a) Reproduce Dr. Bell's reasoning that 
the police can test any one of the guards' statements by questioning in the
manner she suggests. (b) Reproduce Mr.\ Doyle's reasoning that it is 
not possible for his clients to know which statement is being tested unless 
they communicate.
\vspace{3mm}

`Indeed', said Dr. Bell. `If the police accept the offer of asking a single
question   of each of your clients they will wish to make sure that
communication  between them is impossible precisely to ensure that your clients
cannot know which guard's statement is being tested.' However, I think that
there may yet be a way to solve this problem. The theory which describes the
behavior of elementary  particles, called quantum mechanics, has a very strange
property referred to as  {\em non-locality}. While it does not allow
instantaneous communication, it may  nevertheless be sufficient to solve our
problem. I must investigate the question further. How long do we have?'

`Two or three days at the outside, I would say', I replied.

`Excellent!', said Dr.\ Bell. `Then call me mid morning tomorrow, and we shall 
see if I do not have something for you.'

Well I must say that I was highly sceptical. This ``quantum mechanical
non-locality'' to which the Doctor referred sounded to me more like the ravings
of an eccentric than hard science. Perhaps the good Doctor's recent boredom 
had sent her a little over the edge? However, I agreed to call the next
morning, and thanking her for the advice and the cookies, I left for home.

\vspace{2mm}
\vspace{2mm}
\begin{center}
***
\end{center}

Calling Dr.\ Bell the next morning, as I had promised, I found her in
excellent spirits.

`I have good news for you Mr.\ Doyle', she said, `Quantum mechanical
non-locality is indeed sufficient to solve your problem, so long as the police
will agree to ask each guard a single question. Moreover, I have contacted some
colleagues of mine at a laboratory which specializes in quantum information,
and they are able to construct the devices which you will require. I should
have the gadgets in my possession by tomorrow afternoon.'

This was superlative news indeed! I lost no time in making Dr.\ Bell's suggested
counter-offer to the prosecution. To my gratification they accepted it later
that day, and the following afternoon I was back in Dr.\ Bell's office, sitting
in one of her comfortable chairs and tucking into another batch of freshly baked 
cookies. 

`We are lucky, Mr.\ Doyle, that quantum technology is now at the point where we
can manufacture these little gismos.' Dr.\ Bell was holding a small  object,
the size and shape of an electronic car key, and there were two more like it on
the coffee table between us. On each were two buttons, labelled ``lock" and
``unlock".

`These devices contain elementary particles --- in this case electrons --- in a
joint quantum state which is described as being {\em entangled}. Because of
this the results of measurements on individual particles will be correlated.
You must slip these to your clients when you next meet, and explain to them
what to do. Each of your clients is to take one of them with him in his pocket
when he is questioned. If he is asked about the color of the back of his suit,
then he should press the ``lock" button. The device will then vibrate for a few
seconds. If it vibrates constantly, then he should answer ``green''. If it
vibrates in pulses he should answer ``red''. Alternatively, if he is asked
about the color of the front of his suit, then he should press the ``unlock" button, 
and answer depending on the vibration in the same way. This will
guarantee that no matter which guard's testimony is being tested by the police,
it will be confirmed by your clients answers.'

`But without the device's communicating to each other which questions the police 
have asked each of my clients, surely that is impossible!', I replied.

`Not, impossible, Mr.\ Doyle, just very strange', Dr.\ Bell assured me. `Although 
there is no physically detectable signal between
the devices, the quantum particles do influence each other, both at a distance and
apparently instantaneously, '

Having done a physics-for-poets course in my law degree, I was not put off this
easily. `That can't be right. An instantaneous action at a distance would
violate Einstein's theory of relativity which forbids faster-than-light
communication'

Dr.\ Bell smiled enigmatically. `One might think so', she said `but it turns
out that this nonlocal influence cannot under any circumstances be used to
communicate information. You will note that when your clients use the devices, 
none of them will learn what questions the other two have been asked, or what 
their answers are, so no information is communicated between them. Einstein's 
theory survives, although only by the skin of its teeth. This is one of the 
reasons Einstein was never comfortable with quantum mechanics.' 


`That is truly remarkable', I said, pocketing the devices and handing over a
well-earned check. `Well, I cannot thank you enough for your help. You
have indeed solved a problem which I thought to be impossible.' 

`Really the pleasure is mine Mr.\ Doyle. It is delightful to find a real-life 
use for something as curious and apparently arcane as quantum non-locality.'

\vspace{3mm}
\hspace{-3.5mm}{\em Exercise 3:} Explain in detail how Dr.\ Bell's gadgets worked.
\vspace{3mm}

\section*{Answers to the Exercises}

\vspace{3mm}
\hspace{-3.5mm}{\bf Exercise 1}
\vspace{3mm} 

There is, in fact, an elegant way to see that the four statements cannot all be
true by using the properties of multiplication.\cite{Mermin90} Note first that
the guards' statements concern 6 different things, these being the two sides 
(the back and front) of each of the three robbers, and that each guard saw
three of these sides. As Dr. Bell saw, the statements of the first three guards
are equivalent to each of them claiming that ``of the three sides of the
robbers that I saw, an {\em even} number were green'', and the statement of the
fourth guard amounts to ``of the three sides of the robbers that I saw, an {\em
odd} number were green''. Now see what happens if we associate a number with
the front and back of each robber (giving 6 real numbers), making the value 1
if the color is red, and $-1$ if the color is green. Now, since the first three
guards saw an even number of green sides, the product of their three numbers is
plus one, while the product of the three numbers for the forth guard is minus
one. Therefore, the four statements together imply that the product of all of
the guards' numbers (i.e. twelve numbers) is minus one. However, using the
rules of multiplication it is easy to see that this is not possible. If we
examine (using Table~\ref{table}), the twelve various sides that the guards
saw, we see that each of the six different sides appears exactly twice in this
set of twelve. Thus, the product of the associated set of twelve numbers is
actually the product of the squares of the six numbers associated with each
side. Since squares are always positive, this product must be positive. Thus
all four statements cannot be true simultaneously.  

We do not know of a similarly elegant procedure which demonstrates that any
three of the statements alone are consistent, but it is enough to find four
situations which satisfy each of the four subsets of three statements, and this
is not difficult to do by inspection of Table~\ref{table}. In fact, since the
statements of the first three guards are symmetric under an interchange of two
of the robbers, we need only find two situations, one which satisfies the first
three statements, and one which satisfies the last statement along with two of
the first three; interchanging the identities of the robbers will then provide
the others. If all the robbers have red backs and green fronts, then the
statements of the first three guards are true. If robber A is green on the back
and front, and B and C are green on the front and red on the back, then the
statements of guards 2, 3 and 4 are true. Thus any three of the guards could be
telling the truth, but at least one is either mistaken or lying.

\vspace{3mm}
\hspace{-3.5mm}{\bf Exercise 2}
\vspace{3mm}

(a) With only three yes/no questions, the prosecution can only find out the
color of three of the robber's sides. Now, from the discussion in the answer
to Exercise 1, above, we know that each guard's statement concerns only whether
there are an even or odd number of a given color among the three sides that he
saw. As a consequence, to verify any one of the statements the prosecution must
know the colors of {\em all} of the sides which the statement in question
concerns. Thus, since each of the guard's statements concerns a different set
of three sides, the prosecution can only determine the truth or falsity of one
of the statements. 

(b) The reason that Mr.\ Doyle's clients cannot know which statement is being
tested against their answers is as follows. At the time of questioning each
suspect will know only whether he is being asked about the color of his front
or his back. An inspection of Table~\ref{table} shows that for each side
of each robber, there are two of the guards' statements which apply to
it. Thus each suspect will know only that one of these two possible
statements is  being tested. Now, since any two of the statements are
mutually consistent each suspect can choose his answers to agree with those
two statements. However, further examination shows that for each
statement that the prosecution might test against, each suspect will be
trying to satisfy a {\em different} pair of questions. For example, from
Table~\ref{table}, if the prosecution decides to test the statement of
the first guard, then suspect A will know that he is being tested against
either statement 1 or 4, suspect B that it is statement 1 or 3, and suspect C that
it is statement 1 or 2. Thus, together, Mr. Doyle's three clients will be trying
to satisfy {\em all}  four statements. But since the four statements
are inconsistent, they cannot decide beforehand on a set of answers which
will do this.

\vspace{3mm}
\hspace{-3.5mm}{\bf Exercise 3}
\vspace{3mm}
 
In order for Mr.\ Doyle's clients to answer the questions put to them in such a
way that they could guarantee their answers are consistent with the statement
which the prosecution is testing, they would have to determine their answers in
a coordinated fashion. In a universe which obeyed the rules of classical
physics, this would be impossible, because they are prevented from
communicating. However, they are able to achieve this task by using the
following remarkable non-local property which quantum systems possess: It is
possible to prepare two or more quantum systems in a joint state, such that
when the systems are separated (so that communication between them is
impossible), the relationship between the results of measurements made on the
separated systems depends upon {\em what} measurements were made on the
distant systems. It is {\em as if} the quantum systems had been able to communicate
about what measurements were being made on them, and used this information to
arrange the relationship between the measurement outcomes. This effect cannot
be used for communication by three people in possession of the quantum systems,
however, because each person cannot influence {\em which} outcome the others
receive, merely the relationship between all their outcomes. 

Dr.\ Bell's plan was to use the non-locality of quantum mechanics by preparing
three quantum systems in a joint state, giving one to each suspect, and then at
the time of questioning, having each suspect make one of two possible
measurements on his own system depending upon which of the two questions they
were asked. They would then answer their respective questions by using the
result each obtains from his measurement. The joint state that gives precisely
the right answers is the Greenberg-Horne-Zeilinger, or GHZ, state of three
spin-half particles.\cite{GHZ} If you are not familiar with the mathematical
formalism which is used to describe states of, and measurements upon, quantum
systems, then unfortunately at this point you will just have to take our word
for it that the GHZ state, along with suitable measurements, allows the suspects 
to answer the questions so as to cheat the prosecution. However, if you are
familiar with elementary quantum mechanics, then the details of the scheme may
be explained quite simply. 

If we denote the two spin-half eigenstates of the operator for spin in the $z$
direction as $\left|\uparrow\right\rangle$ for ``spin up'' and
$\left|\downarrow\right\rangle$ for ``spin down'', then the GHZ state is
\begin{equation}
 \left|\mbox{GHZ}\right\rangle = \frac{1}{\sqrt{2}} (\left|\uparrow\right\rangle_{\rm A} \left|\uparrow\right\rangle_{\rm B} \left|\uparrow\right\rangle_{\rm C} - \left|\downarrow\right\rangle_{\rm A} \left|\downarrow\right\rangle_{\rm B} \left|\downarrow\right\rangle_{\rm C}) .
\end{equation}
Here the subscripts indicate which system belongs to which suspect (A, B, or C). 
Dr.\ Bell's gadgets work as follows: If a suspect is asked what color his suit
was on the front, then he presses the ``lock" button. This  triggers a measurement
which projects his system onto one of the basis states
$\{\left|\otimes\right\rangle,\left|\odot\right\rangle\}$, where these states
are given by
\begin{eqnarray}
   \left|\otimes\right\rangle & = & \frac{1}{\sqrt{2}} ( \left|\uparrow\right\rangle + i\left|\downarrow\right\rangle ) , \\
   \left|\odot\right\rangle & = & \frac{1}{\sqrt{2}} ( \left|\uparrow\right\rangle - i\left|\downarrow\right\rangle ) .\label{crossstate}
\end{eqnarray}
This corresponds to a measurement of the spin of the particle in the
$y$-direction. If the gadget gets the result corresponding to
$\left|\otimes\right\rangle$ then it vibrates in a way that tells him to answer
that the color was red. Similarly, if the result is
$\left|\odot\right\rangle$ then he will know to answer that the color was green.
Alternatively, if a suspect is asked what color his suit was on the back,  he
presses the ``unlock" button and this makes a measurement which projects the system
onto one of the states $\{\left|
\rightarrow\right\rangle,\left|\leftarrow\right\rangle\}$, where
\begin{eqnarray}
   \left| \rightarrow\right\rangle & = & \frac{1}{\sqrt{2}} ( \left|\uparrow\right\rangle + \left|\downarrow\right\rangle ) , \\
   \left|\leftarrow\right\rangle & = & \frac{1}{\sqrt{2}} ( \left|\uparrow\right\rangle - \left|\downarrow\right\rangle ) .
\end{eqnarray}
This is a measurement of the spin of the particle in the $x$-direction. If the
suspect gets the result corresponding to $\left| \rightarrow\right\rangle$ then
he answers that the color was red, otherwise he says that it was green. A
quantum mechanical analysis of the two measurements shows that every set of
possible outcomes of three of these measurements is consistent with all the
statements of the guards, and in particular will be consistent with any
statement  that the police choose to test by asking their three questions. For
example, assume the police are checking if the suspects' answers are consistent
with the first guard's statements. Then suspect A will be asked about the back 
of his suit, so he will make an $x$ spin measurement. The result of the
measurement is either $\left| \rightarrow\right\rangle$ or
$\left|\leftarrow\right\rangle$, and let us assume that it is $\left|
\rightarrow\right\rangle$ or red. In this case the action of the measurement
is to apply the projection operator  $\left|\rightarrow\right\rangle
\left\langle\rightarrow\right|$ to A's system, and  renormalise the state
(which in this case involves multiplying by $\sqrt{2}$). The state of the
three systems after the measurement is then  
\begin{eqnarray}
 \sqrt{2} (\left|\rightarrow\right\rangle \left\langle\rightarrow \right| )_{\rm{A}}\left|\mbox{GHZ}\right\rangle 
   & = & \left|\rightarrow\right\rangle_{\rm A} ( \left\langle\uparrow\right| + \left\langle\downarrow\right|)_{\rm A}  
   \left|\mbox{GHZ}\right\rangle \nonumber \\
   & = & \left|\rightarrow\right\rangle_{\rm A} \left( \frac{\left|\uparrow\right\rangle_{\rm B} \left|\uparrow\right\rangle_{\rm C} -  \left|\downarrow\right\rangle_{\rm B} \left|\downarrow\right\rangle_{\rm C}}{\sqrt{2}}\right) \nonumber \\
   & \equiv & \left|\rightarrow\right\rangle_{\rm A} \left|\mbox{BC}\right\rangle
 \label{BCstate}
\end{eqnarray}
Now suspect B is asked about the front of his suit, so he makes a $y$ spin
measurement. The result can be either $\left|\otimes\right\rangle$ or
$\left|\odot\right\rangle$, but let's assume the result is
$\left|\otimes\right\rangle$ or red. The state in Eq.~(\ref{BCstate}) now becomes  
\begin{eqnarray}
 \sqrt{2} ( \left|\otimes \right\rangle \left\langle\otimes \right| )_{\rm{B}}\! \left|\rightarrow\right\rangle_{\rm A}\! \left|\mbox{BC}\right\rangle 
 & = &  \left|\otimes \right\rangle_{\rm{B}}\! ( \left\langle\uparrow\right| -i \left\langle\downarrow\right|)_{\rm B}\! \left|\rightarrow\right\rangle_{\rm A}\! \left|\mbox{BC}\right\rangle \nonumber \\
 & = & \left|\rightarrow\right\rangle_{\rm A} \left|\otimes \right\rangle_{\rm{B}} \left( \frac{ \left|\uparrow\right\rangle_{\rm C} + i \left|\downarrow\right\rangle_{\rm C} }{\sqrt{2}} \right) \nonumber \\
 & = &\left|\rightarrow\right\rangle_{\rm A} \left|\otimes \right\rangle_{\rm{B}} \left|\otimes \right\rangle_{\rm{C}} ,
 \label{Cstate}
\end{eqnarray} 
(Note the minus sign appearing in $\left\langle\otimes \right|_{\rm{B}}$
because $ \left\langle\otimes \right|_{\rm{B}}$ is the adjoint of
$\left|\otimes\right\rangle_{\rm{B}}$.) Since C's system is now  in the state
$\left|\otimes \right\rangle$, when C is asked about the front of his suit his
measurement must give $\left|\otimes\right\rangle$ or red. Thus, the police
find that there are no green suits among the answers, which is consistent with
the first guard's testimony that he saw an even number of green suits. One
could try out all the other possibilities and one would find that in each case
the suspects' answers will be consistent with the testimony of the guard chosen
by the police. Also, notice that the order in which the suspects give their
answers does not matter. There is an elegant treatment of this problem in
Mermin's paper,\cite{Mermin90} where this three-particle GHZ-style proof of
Bell's theorem was first presented.

The gadgets provided by Dr.\ Bell do not yet exist. However, spin-based
qubits\cite{spinqbits} are one of the contenders for scalable quantum
information processing. With the current rapid advances in quantum information
technology,\cite{QIC:IQC} there is no reason to assume that devices operating
as we describe could not be built within the next few decades.

\section*{Acknowledgments}
We thank Damian Pope for helpful discussions. We note also that the basic structure
of Figure~\ref{map} was inspired by the diagram illustrating the GHZ result in
Ref.\cite{Maudlin}


\begin{thebibliography}{99}
\bibitem{footnote} 
   In the first case one must know the no-cloning theorem, and in the second one 
   must know that the fact that a qubit can only contain one bit of information, 
   despite being preparable in infinitely many different ways.
\bibitem{footnote2} 
For readers who are familiar  with Mermin's illustration of
nonlocality~\cite{Mermin90}, upon which the situation here is based, it may be
helpful to note which elements of our scenario correspond to those of Mermin's.
The three clients in our  story take the place of Mermin's three detectors, and
the two settings on these  detectors to the two sides of each of the robbers
(setting 1 to the back  and setting 2 to the front). The four different
combinations of detector settings therefore correspond to the four different
combinations of sides of the robbers seen by the four guards. Finally, the two
colours that the detectors can flash correspond to the same two colors of the
robbers' suits. Thus, in Mermin's case when he sets the detector settings to
111, for example, and states that an even number of the three detectors flash
red, that corresponds in our case to saying that the fourth guard saw an even
number of the robbers wearing red.
\end{thebibliography}
\end{document}